\documentclass[3p,twocolumn]{elsarticle}

\usepackage{graphicx}
\usepackage{upgreek}
\usepackage{units}
\usepackage{amsmath}
\usepackage{amssymb}
\usepackage[latin1]{inputenc}
\usepackage{xspace}
\usepackage{array}
\usepackage{booktabs}

\newcommand{\MeVc}{\ensuremath{$MeV\!/c$ }\xspace}

\newcommand{\mmmrad}{\,mm\thinspace\thinspace mrad\xspace}
\newcommand{\degree}{\ensuremath{^\circ}\xspace}
\newcommand{\um}{\ensuremath{\upmu}m}

\begin{document}

\begin{frontmatter}
\title{Reflective Optical System for Time-Resolved Electron Bunch Measurements at PITZ}
\author{K.~Rosbach\corref{cor1}}
\ead{Kilian.Rosbach@cern.ch}
\author{J.~Bähr}
\address{DESY, Platanenallee 6, 15738 Zeuthen, Germany}

\author{J.~Rönsch-Schulenburg}
\address{Hamburg University, Luruper Chaussee 149, 22761 Hamburg, Germany}

\cortext[cor1]{Corresponding author, now at CERN, Mailbox J23810, 1211 Meyrin 23, Switzerland.}

\begin{abstract}
The Photo-Injector Test facility at DESY, Zeuthen site (PITZ), produces pulsed electron beams with low transverse emittance and is equipped with diagnostic devices for measuring various electron bunch properties, including the longitudinal and transverse electron phase space distributions.
The longitudinal bunch structure is recorded using a streak camera located outside the accelerator tunnel, connected to the diagnostics in the beam-line stations by an optical system of about~\unit[30]{m} length. This system mainly consists of telescopes of achromatic lenses, which transport the light pulses and image them onto the entrance slit of the streak camera. Due to dispersion in the lenses, the temporal resolution degrades during transport. This article presents general considerations for time-resolving optical systems as well as simulations and measurements of specific candidate systems. It then describes the development of an imaging system based on mirror telescopes which will improve the temporal resolution, with an emphasis on off-axis parabolic mirror systems working at unit magnification. A hybrid system of lenses and mirrors will serve as a proof of principle.
\end{abstract}

\end{frontmatter}

\textsc{Keywords}: photo-injector, longitudinal phase space diagnostics, streak camera, time-resolved imaging, Cherenkov imaging, reflective optics

\section*{Abbreviations used in this article}
\begin{tabular}{ll}
DESY & Deutsches Elektronen-Synchrotron\\
FLASH & Free electron LASer Hamburg\\
IS & Input Section\\
MTF & Modulation Transfer Function\\
OAP & Off-Axis Parabolic\\
OTL & Optical Transmission Line\\
PITZ & Photo-Injector Test facility at Zeuthen\\
XFEL & X-ray Free Electron Laser\\
\end{tabular}

\section{Introduction}
\label{sec:introduction}
The Photo-Injector Test facility DESY, Zeuthen site (PITZ), is a linear electron accelerator~\citep{PITZ_pro99}. It produces electron bunches with a nominal charge of~\unit[1]{nC} by the photoelectric effect, using UV laser pulses with a flat-top temporal distribution of $\sim\!\!\unit[20]{ps}$ FWHM duration. The gun cavity accelerates the electron bunches to a momentum of about~\unit[6.5]{\MeVc}, which is increased to about~\unit[15]{\MeVc} in a booster cavity~\citep{Stephan2010}. Under these conditions a projected transverse emittance of~\unit[1.0]{\mmmrad} was measured~\citep{IPAC2010_SR}. Several gun cavities of similar designs (with variations in cooling systems and surface treatments) were conditioned and characterized at PITZ, and delivered to the Free Electron Laser in Hamburg (FLASH,~\citep{FLASH_waterwindow_2007}). The main research goals for PITZ are the optimization of electron bunch parameters for driving short wavelength free electron lasers, in particular reaching a transverse emittance suitable for the European X-ray Free Electron Laser (XFEL, \citep{XFEL}), as well as extensive R\&D on electron source design and diagnostics.
\newpage
The temporal bunch structure is recorded by a Hama\-matsu 
C5680 streak camera, which provides a temporal resolution better than~\unit[2]{ps} in the visible spectrum~\citep{StreakManual}. This sensitive device was placed in a separate room outside the accelerator tunnel to protect it from radiation and to allow camera handling during machine operation. Currently, four read-out ports along the electron beam-line are connected to the streak camera by an optical transmission line (OTL) of about \unit[30]{m} length \citep{DIPAC2003_JB, DIPAC2007_JB}. At every read-out port, a radiator can be moved into the beam tube. When passing the radiator, the electron bunches create light pulses, either as Cherenkov light (typically using silica aerogel as radiator material) or as optical transition radiation (using, for example, silicium plates with an aluminum coating). The transverse and longitudinal structures of the light pulse produced by the radiator closely resemble those of the original electron bunch~\citep{Baehr2004_Aerogel}. A significant part of the emitted light is contained in the visible spectrum and can be transported by conventional optical elements.

\begin{figure*}[t!]
  \centering
  \includegraphics[width=16cm]{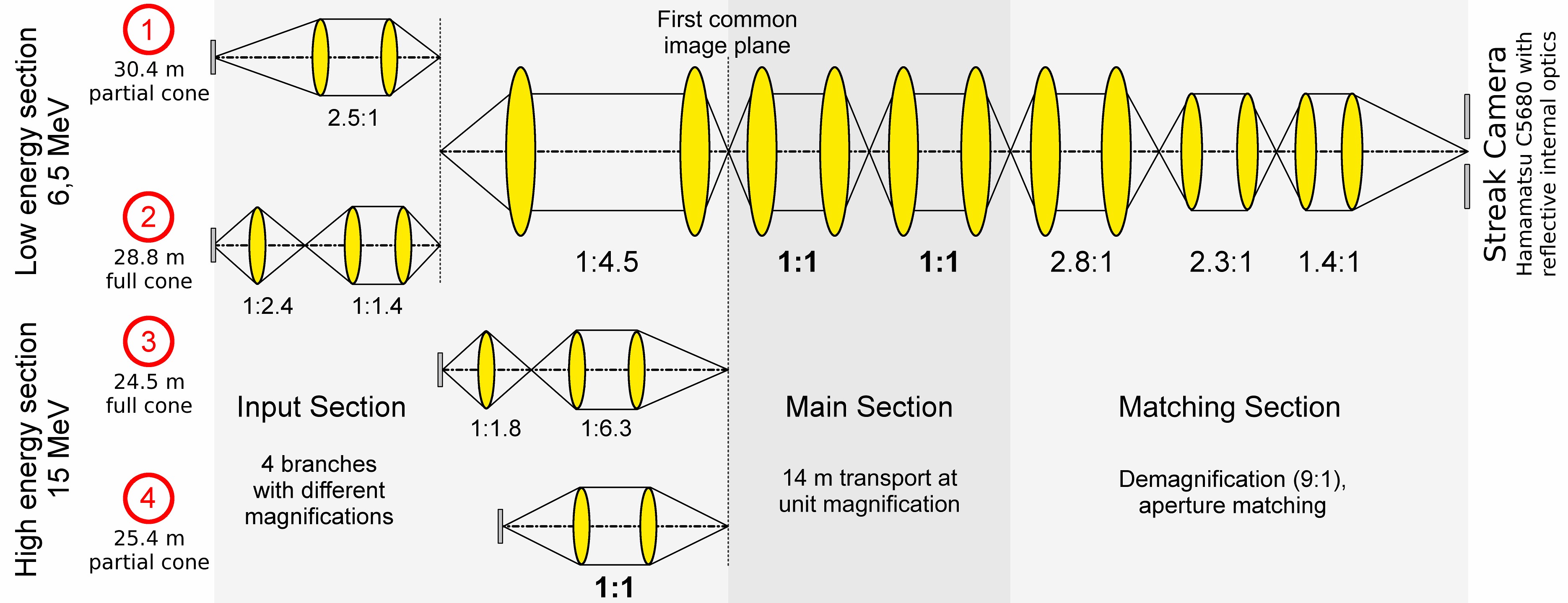}
  \caption{Simplified scheme of the current optical transmission line (OTL), not to scale. Only the most commonly used elements are included for clarity. The dashed lines indicate positions of movable mirrors that allow switching between read-out ports. There are three subsystems with unit magnification (1:1). The branches labeled ``partial cone'' use only a fraction of the Cherenkov cone for imaging (see figure~\ref{fig:partial_cone}).}
  \label{fig:overall_system_layout}
\end{figure*}

The currently installed optical system (figure~\ref{fig:overall_system_layout}) is comprised of about one dozen lenses per branch and can be used to measure the electron bunch length~\citep{FEL2003_JB} and the electron distribution in the longitudinal phase space span\-ned by longitudinal position~$z$ and longitudinal momentum~$p_z$ \citep{FEL2005_JR}. However, the temporal resolution deteriorates due to dispersion in the lenses, since the glass introduces a wavelength-dependent delay. Most of the lenses in the system are achromats, which correct the resulting chromatic aberration, but the effect on the temporal structure cannot be compensated in this way. The temporal resolution can be restored to an acceptable level by inserting an interference filter of narrow bandwidth (typically: \unit[10]{nm}~FWHM transmittance, centered at 500 or \unit[550]{nm}) into the optical path, but the loss of light at the filter leads to an intensity reduction by about 96\%. This necessitates an integration over several pulses to reach an adequate signal-to-noise ratio, and the temporal structure of an individual bunch cannot be recorded. Furthermore, the lenses are sensitive to the radiation in the accelerator tunnel. If the radiation is too strong, the lenses become brown and thus less transmissive, leading to a further reduction of recorded intensity at the streak camera~\citep{NIM10_JR}.
This article presents an investigation toward a new optical system based on mirrors. Since the law of reflection is not wavelength-dependent, the temporal resolution will not be diluted by dispersion. Also, mirrors do not suffer from the radiation in the tunnel: Several plane mirrors of borosilicate glass with a reflectivity coating are present in the current optical system, and no discoloration was observed for them. Therefore, a reflective optical system intrinsically avoids the loss of temporal resolution due to dispersion and is resistant to radiation damages. On the other hand one has to deal with a more challenging geometry and a more difficult adjustment.

\section{General Considerations for Reflective Optics}
\label{sec:gener-cons-refl}
The longitudinal electron phase space distribution contains the temporal structure of the bunch as well as its momentum distribution. At PITZ, it is measured using a dipole spectrometer in conjunction with a streak camera~\citep{Lipka}. The optical system between the radiator in the spectrometer and the streak camera has to maintain the spatial and temporal information of the light pulses.
Fortunately, spatial and temporal resolution are closely related for optical systems which operate independently of wavelength, as can be seen from Fermat's principle:
From all the possible paths connecting two points~$\mathcal A$ and~$\mathcal B$, any optical path~$\mathcal P$ realized in nature is an extremum of the optical path length~$\mathcal L$. On the other hand, $\mathcal L$ is directly proportional to~$t$, the traveling time of the ray along~$\mathcal P$:
\begin{eqnarray}
  \mathcal L&=&\underset{\mathcal P}\int n\left(\vec s\right)\mbox d \vec s,\\
  t&=&\underset{\mathcal P}\int \frac1{c_n\left(\vec s\right)}\mbox d \vec s=\frac1{c_0}\mathcal L,
  \label{eq:def_optical_path_lengh}
\end{eqnarray}
where $\vec s$ is an element of the path $\mathcal P$, and $c_0$ and $c_n$ are the speed of light in vacuum and in a medium with refractive index $n$, respectively. Assuming that all physical optical paths connecting~$\mathcal A$ to~$\mathcal B$ are related by a continuous transformation parametrized by~$\alpha$, then~$t$ is constant for all of these paths, since
\begin{eqnarray}
  \frac{\mbox d}{\mbox d\alpha}t=\frac1{c_0}\frac{\mbox d}{\mbox d\alpha}\mathcal L=0,
  \label{eq:fermats_principle}
\end{eqnarray}
where in the final step Fermat's principle is used again. This means that a system with a perfect spatial resolution, i.e. a system in which all the rays originating from~$\mathcal A$ are arriving at~$\mathcal B$, will also have a perfect temporal resolution due to identical traveling times of all rays. The argument is true only for the two points in consideration and might not hold for an arbitrary pair of points in the vicinity of $\mathcal A$ and $\mathcal B$, respectively. Typically this is improved by demanding that the derivatives of~$\mathcal L$ with respect to lateral displacements of~$\mathcal A$ and $\mathcal B$ vanish, which then leads to the Abbe sine condition~\citep{Korsch}. Since this condition is often met in well-corrected optical systems, a mirror system with a good spatial resolution is likely to exhibit a good temporal resolution. This provides suitable starting points for the following investigation. It is important to notice that the whole argument does not hold for lenses or any other system where dispersion cannot be neglected. In those systems, the extremal path is a function of the wavelength -- but the extremum of Fermat's principle is found with respect to a variation of the path for a fixed wavelength (as opposed to a variation of wavelength itself).

In the experimental setup the optical pulse first has to pass a quartz window enclosing the Silica aerogel inside the vacuum system and then leaves the beam tube vacuum through another quartz window, before traveling in air. Windows and air introduce dispersion, which results in a traveling time delay between components of different wavelengths. This delay can directly be calculated from the refractive indices and thicknesses and the corresponding group velocity. The refractive index of air was estimated using the modified \'Edlen equation~\citep{BirchDowns1,BirchDowns2}. For a light pulse on the optical axis, the difference in traveling time between the \unit[700]{nm} component and the more strongly delayed component at~\unit[400]{nm} is~\unit[2.12]{ps} for~\unit[30]{m} air at normal conditions, and~\unit[1.30]{ps} for~\unit[8]{mm} quartz. The net contributing wavelength range is determined by the Cherenkov intensity spectrum and the transmittivity/reflectivity of the optical elements. If necessary, it can be further restricted with an interference filter. For a wavelength range of, for example, $500$--$\unit[600]{nm}$, the aforementioned delays are reduced to~\unit[0.60]{ps} and~\unit[0.36]{ps}, respectively.

In summary, there are two factors degrading temporal resolution: Traveling time differences due to dispersion and due to geometry. The estimate given above for the dispersion introduced by air and by the quartz windows, which will also affect an all-reflective optical system, already exceeds the streak camera resolution of about \unit[2]{ps}~FWHM if the full visible spectrum is used. An interference filter will probably still be necessary, but it can have a much wider wavelength acceptance. The delay due to geometry depends on the design of the optical system to be discussed in the following sections. It cannot be reduced by an interference filter.

\section{Different Sections of the Optical System}
\label{sec:diff-sect-optic}
The optical system at PITZ consists of three sections: the input section (IS), the main section and the final matching section (see also figure~\ref{fig:overall_system_layout}). There are four read-out ports, each with its own input optics, referred to as ``input branches''. At the beginning of each branch, light pulses are created when the electron beam passes through a Silica aerogel with refractive index~$n_{Aerogel}$. The light pulses are initially strongly divergent, with different opening angles~$\vartheta_{in}$. The transport at large opening angles requires that the optical elements have large diameters or are located at small distances. Both options are impractical, so instead one applies a lateral magnification by a factor $m_{IS}$ in each of the branches of the input system. This reduces the tangens of the opening half angle, $\tan\vartheta$, by the same factor. The values of $n_{Aerogel}$, $\vartheta_{in}$, $\vartheta_{out}$ and $m_{IS}$ for the different branches of the current OTL are summarized in table~\ref{tab:input_sections}.

\begin{figure}[htbp]
  \centering
  \includegraphics[width=7.5cm]{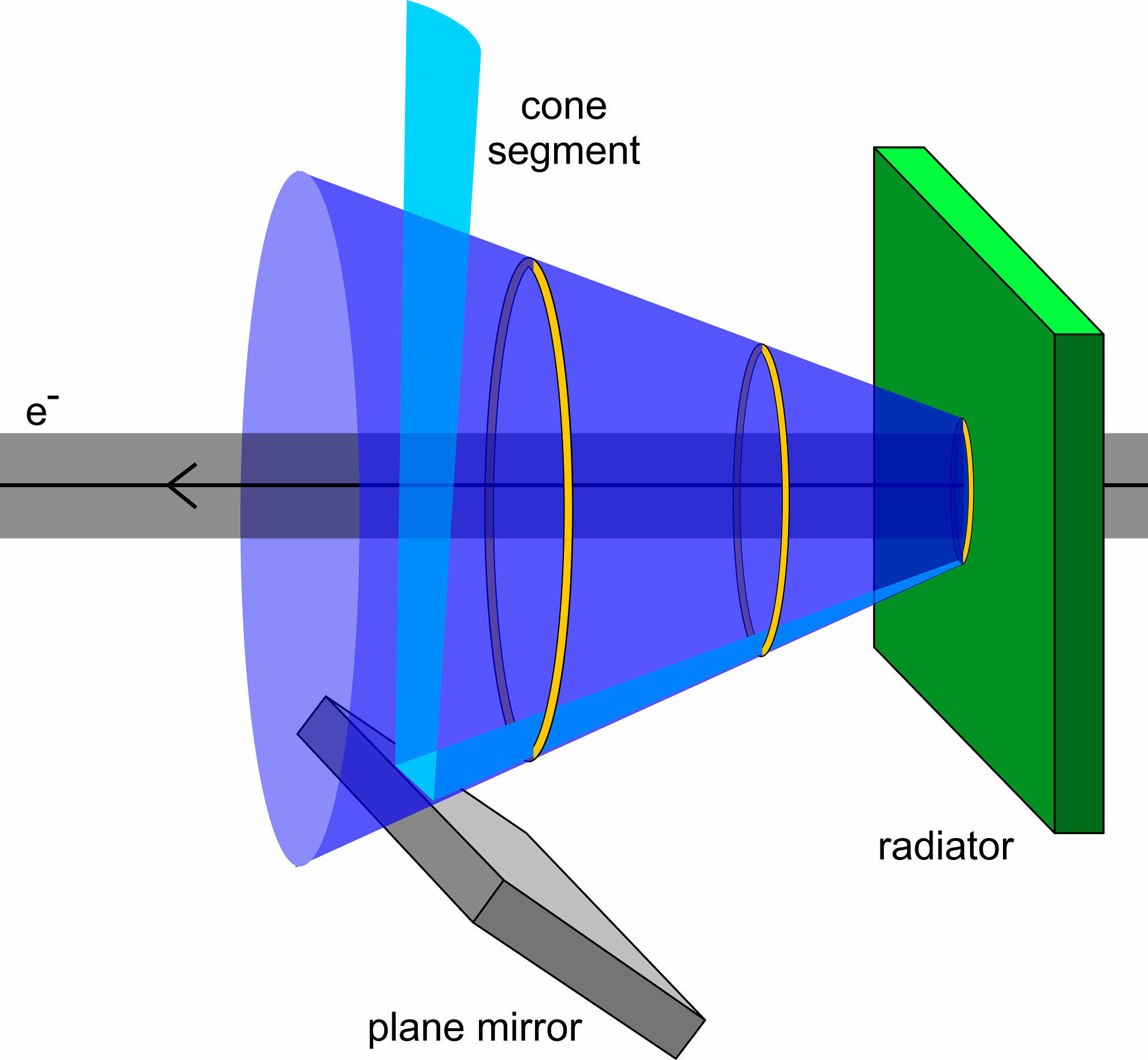}
  \caption{A plane mirror can be used to slice out a segment of the Cherenkov cone. This setup is referred to as "partial cone".}
  \label{fig:partial_cone}
\end{figure}

The choice of the radiator materials and $m_{IS}$ is motivated by the different applications: Branches 1 and 4 are used for measurements of the longitudinal phase space distribution and have to transport extended objects (several~\unit{cm$^2$}) created by the dipole spectrometer. The aerogel has to produce a sufficient light intensity and thus the index of refraction must not be chosen too low. The index of $n_{Aerogel}\!=\!1.05$ produces Cherenkov light with large cone angles (18\degree-19\degree, depending on the electron energy), which cannot be fully collected. Figure~\ref{fig:partial_cone} shows the so-called ``partial cone'' setup~\cite{DIPAC2003_JB}, which is used to slice out a small, almost-parallel segment of the cone. Branches 2 and 3 are used to record just the temporal bunch structure. The transverse electron bunch profile extends only over a small area (few \unit{mm$^2$}) and consequently has a higher intensity there. A lower index of refraction will still produce a sufficient light intensity, while resulting in small opening angles of a few degree that can be fully collected and transported. Finally, $m_{IS}$ has to be chosen such that the images created by the input sections all can be transported by the main section common to all branches. Movable plane mirrors allow to switch between different input branches.

The main section of the system covers most of the remaining distance. The image  is transported at unit magnification. This is favorable because it allows designing systems with a high degree of symmetry, which usually have a better cancellation of aberrations and a higher number of identical elements, improving imaging quality while reducing production cost. In the final section of the system a lateral demagnification is used to match the image to the size of the streak camera slit and the aperture of the input optics. The main section, the final matching optics, and the streak camera slit are common to all input branches.

\begin{table}[b]
  \centering
  \begin{tabular}{c@{}cr@{.}l@{}r@{}r@{}r@{}r@{}r}
  \toprule
     \multicolumn{4}{c}{} &
     \multicolumn{2}{c}{before IS} & &
     \multicolumn{2}{c}{after IS} \\
   \cmidrule{5-6} \cmidrule{8-9}
     & \multicolumn{3}{c}{$n_{Aerogel}$} &
     \multicolumn{1}{c}{size} &
     \multicolumn{1}{c}{$\vartheta_{in}$} &
     \multicolumn{1}{c}{$m_{IS}$} &
     \multicolumn{1}{c}{size} &
     \multicolumn{1}{c}{$\vartheta_{out}$} \\
   \midrule
1 &&1&05 &40 mm &\,\,\,$\sim$18\degree& 1.8 &       72 mm &        (PC)       \\
2 &&1&008&10 mm &         5.7\degree & 15.0 &\,\,\,150 mm &\,\,\,\,0.38\degree\\
3 &&1&008&10 mm &         7.0\degree & 11.0 &      110 mm &        0.63\degree\\
4 &&1&05 &76 mm &    $\sim$19\degree &  1.0 &       76 mm &        (PC)       \\
    \bottomrule
  \end{tabular}
  \caption{Properties of the input section (IS) for branches 1-4. The "before IS" sizes refer to the sizes of the Cherenkov radiators -- typical beam distribution sizes are smaller. Branches 1 and 4 are used for longitudinal phase space distribution measurements, where a large screen is required and the light has to be collected with the partial cone (PC) setup (see figure~\ref{fig:partial_cone}). Branches 2 and 3 are used only for measurements of the temporal structure of the bunch profile.}
  \label{tab:input_sections}
\end{table}

\section{Goal Specification and Methods}
\label{sec:goal-specification}
Using the overall system layout presented above it is straightforward to compile more specific requirements for the individual sections. The requirements for the main section are summarized in table~\ref{tab:goal_specs} and explained further in the following paragraphs.

The streak camera C5680 has an entrance slit of $\unit[5.4]{mm}$ $\times\unit[0.1]{mm}$ with an aperture of $f/D=4.0$, using reflective input optics. The spatial resolution is \unit[25]{lp/mm}, and the temporal resolution is better than \unit[2]{ps} in the visible range. The slit width is adjustable, but typically $\le\!\unit[100]{\um}$ are used.

Working backward from the streak camera specifications and selecting a magnification of $m_f=0.1$ for the final matching section, one finds that the main section must be able to transport images with a diameter of up to \unit[54]{mm} at an aperture of $f/D\!=\!40$ (corresponding to an opening half-angle of about 0.7\degree), and deliver an intermediate image with a resolution of better than \unit[2.5]{lp/mm} to the final matching section.
This choice of $m_f$ is based on the current OTL. The main section is comprised of several telescopes, which individually have to provide a better resolution.
A resolution of \unit[10]{lp/mm} was chosen as the goal specification for a single telescope in the main section.

The streak camera resolves only one spatial direction, the other one represents the temporal axis. In this latter direction the entrance slit is very narrow (\unit[100]{\um}, see above) because the spatial distribution of the incoming light in this direction will be folded with the temporal distribution and limit the achievable temporal resolution~\citep{ThesisJuliane}. Thus in principle only a thin slice of the image needs to be transported, but for practical reasons a circular intermediate image with a diameter of \unit[54]{mm} at the beginning of the main section was assumed for the following studies.
The overall degradation of temporal resolution in the OTL should be smaller than the streak camera resolution of~\unit[2]{ps}. Considering the effects of vacuum windows and air estimated in section~\ref{sec:gener-cons-refl}, remaining below~\unit[1]{ps} is desirable. For the comparison of different systems, the temporal resolution was approximated by the RMS of the distribution of distances traversed by rays traced through the system. For the hybrid system presented later in this article, the group velocity was also taken into account.

The loss of temporal resolution will not be distributed equally along the system: Preserving the temporal structure is in general more difficult when a magnification or demagnification of the image is involved, compared to transporting the image at unit magnification. This again is a result of the lower symmetry of such systems.

Each mirror surface introduces a folding of the light path. It is important to aim for a high ``length efficiency'', i.e. the ratio of the geometrical distance from object to image and the (average) distance actually covered by the rays passing through the system. For some setups, mirror surfaces will cast shadows on each other, creating an obstruction of the light path. This is described by the ``transmission efficiency'', defined as the fraction of rays emitted from an extended object that can traverse the system without missing a mirror surface or being blocked by an obstruction. It is a purely geometrical quantity and for a given object size it can be estimated by raytracing. In addition to geometrical light-losses, the system will also have losses due to scattering and absorption on the mirror surfaces. These can be reduced by choosing an appropriate mirror coating. At the design level, it is important to aim for a low number of mirror surfaces per unit distance.

Care was taken to understand possible effects of the parallel cone structure of the Cherenkov light. Often this structure leads to a more homogeneous simulated resolution across the object field, but will require larger mirror surfaces to keep light losses at an acceptable level. The Cassegrain telescope mentioned below has a limited angular acceptance and is unsuitable for Cherenkov light imaging.

\begin{table}[b]
  \centering
  \begin{tabular}{lc}
    \toprule
    Transportable image size & \unit[54]{mm} \\
    Aperture & $f/D\!=\!40$ \\
    Optical resolution (in image plane) & \unit[2.5]{lp/mm} \\
    Optical resolution (per telescope) & \unit[10]{lp/mm} \\
    Temporal resolution & $\lesssim\unit[1]{ps}$ \\
    Length efficiency & $>0.9$ \\
    Transmission efficiency & $>0.9$ \\
    \bottomrule
  \end{tabular}
  \caption{Summary of goal specifications for the reflective main section of the optical transmission line.}
  \label{tab:goal_specs}
\end{table}

\section{Mirror Systems with Unit Magnification}
\label{sec:mirror-systems-with}
\begin{figure*}[tb]
  \centering
  \includegraphics[width=16cm]{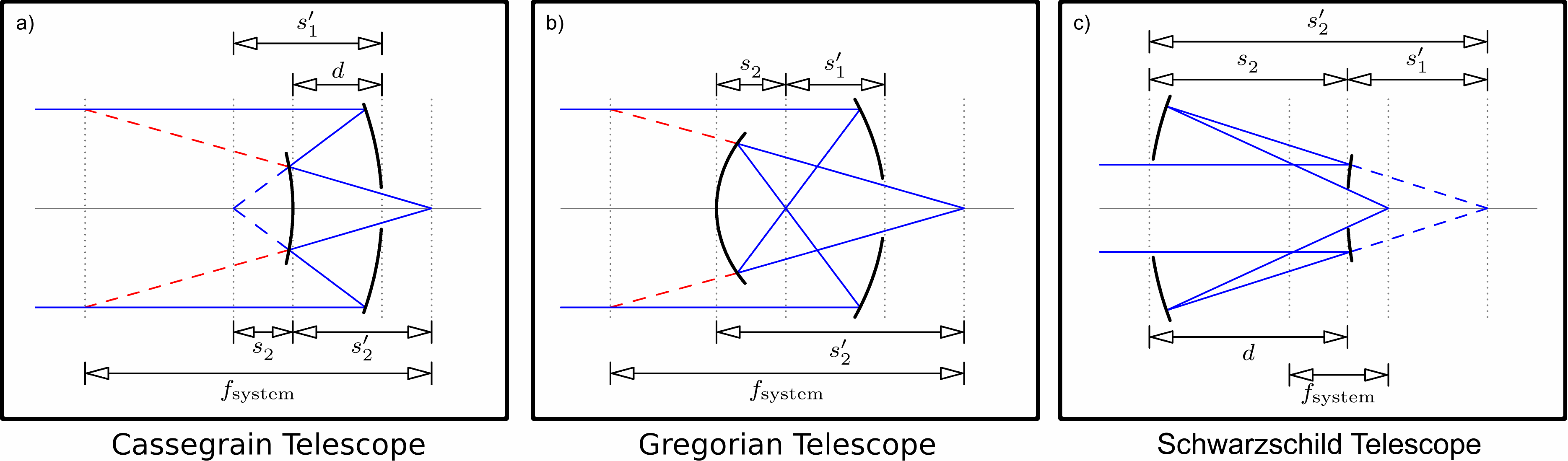}
  \caption{Classical telescopes: a) Cassegrain geometry with a parabolic primary mirror and a hyperbolic secondary mirror, b) Gregorian geometry with a parabolic primary and an ellipsoid secondary, c) Schwarz\-schild geometry with two different oblate ellipsoid mirrors. $s_i$ and $s_i'$ indicate object and image distances of mirror $i$, $d$ is the mirror spacing, and $f_{system}$ is the system focal length. The free parameters were optimized with respect to the design goals, but none of these setups performed satisfactory in simulation.}
  \label{fig:classical_telescopes}
\end{figure*}

Different system designs were considered and numerically optimized with respect to the specified criteria. Most of them were inspired by~\cite{Korsch}, which offers a systematic overview of reflective optical systems and design principles. Once a successful candidate is found, further simulations are needed to confirm that the requirements on manufacturing and alignment precision are reasonable.

Figure~\ref{fig:classical_telescopes} shows systems we studied that are based on classical telescopes (Cassegrain, Gregorian, Schwarz\-schild). However, even after numerical optimization of the system parameters they were found unsuitable because of insufficient light acceptance at small angles (Cassegrain), insufficient temporal resolution (Gregorian), or both (Schwarz\-schild). The latter result was particularly unexpected, since the Schwarz\-schild telescope is fully corrected for all third order (Seidel) aberrations except distortion.

\begin{figure}[tbhp]
  \centering
  \includegraphics[width=7.5cm]{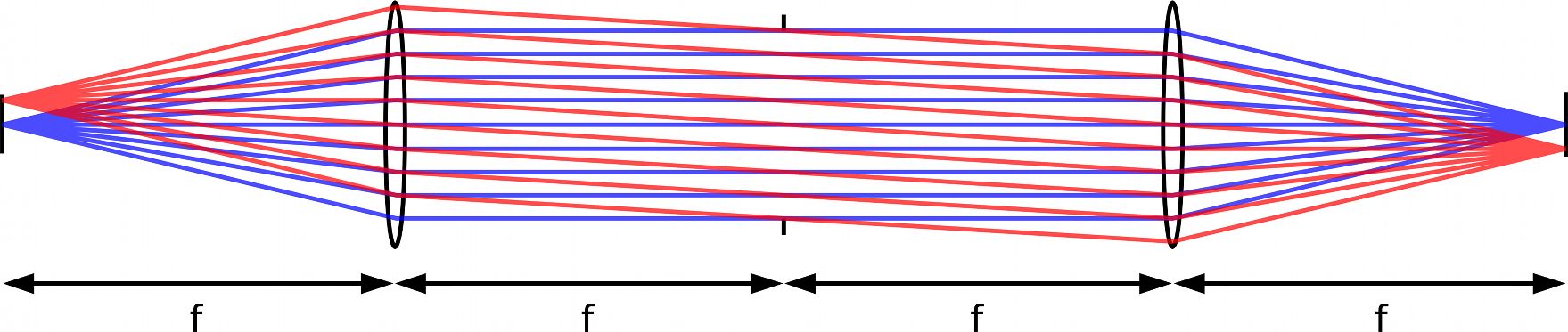}
  \caption{Illustration of the $4f$--scheme, which is used in the current OTL, and will also be used for the new transport system based on mirrors.}
  \label{fig:4fscheme}
\end{figure}

\begin{figure}[tbhp]
  \centering
  \includegraphics[width=7.5cm]{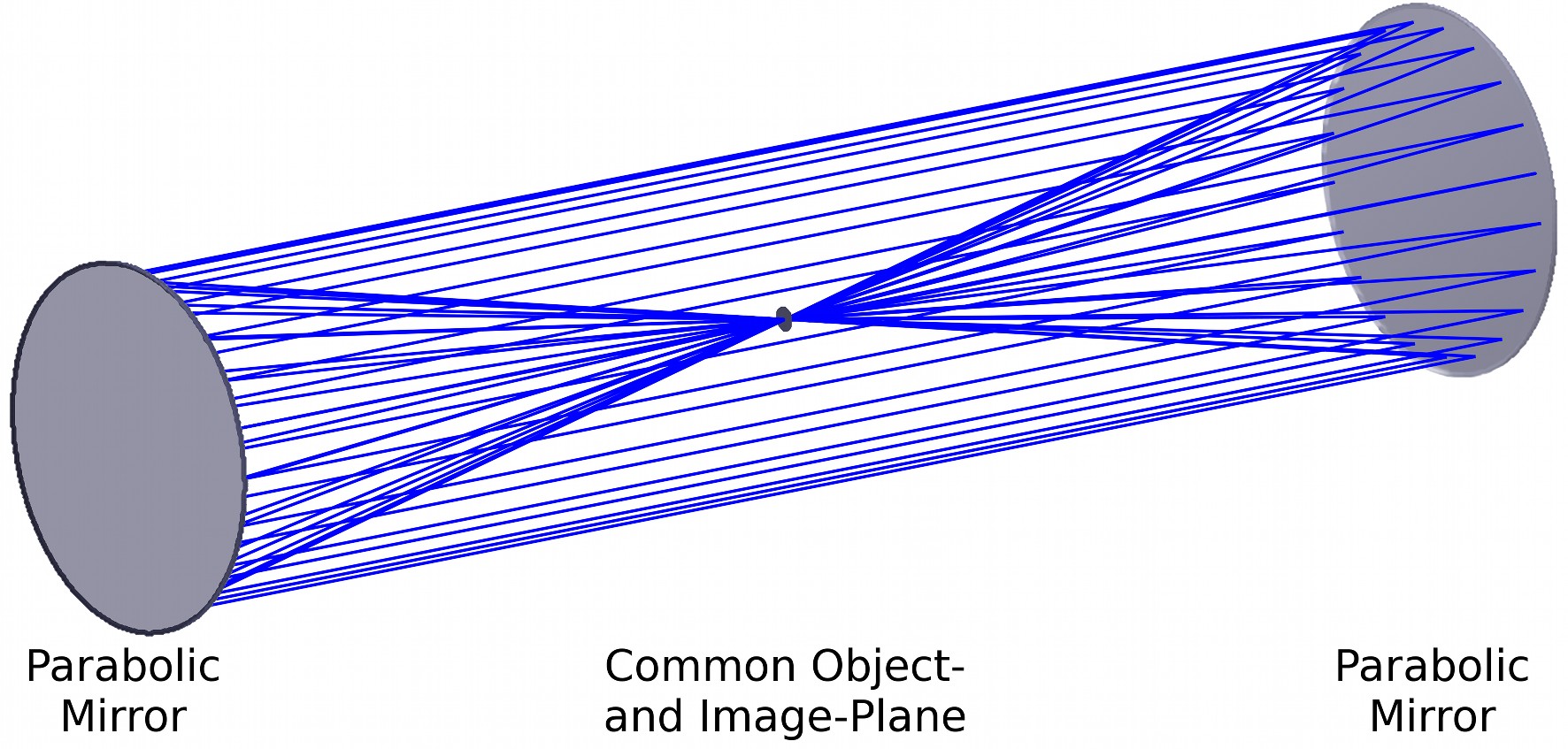}
  \caption{An implementation of the $4f$-scheme with two identical on-axis parabolic mirrors, arranged face to face at a distance of $2f$. This system performs well in terms of spatial and temporal resolution, but has zero length efficiency.}
  \label{fig:parabs_linear}
\end{figure}

The majority of the subsystems in the current OTL follow the ''$4f$--scheme'', i.e. they are comprised of two identical subunits with focal length $f$ that are oriented oppositely and have a distance of $2f$ between their principal planes (figure~\ref{fig:4fscheme}). For a typical lens system, the spacing between two intermediate images is then $4f$, while for a mirror system the distance is smaller. Figures~\ref{fig:4fscheme} and~\ref{fig:parabs_linear} show examples of a lens system and a mirror system, respectively, both complying to the $4f$--scheme.

\subsection{On-Axis Parabolic Mirrors}
\label{sec:on-axis-parab-mirr}
A simple setup of two on-axis parabolic mirrors each with focal length $f$ at a distance of $2f$, arranged face to face on their symmetry axis (figure~\ref{fig:parabs_linear}), achieves a high temporal and spatial resolution over a wide range of choices for~$f$ and mirror diameter~$D$. For such a setup with $f\!=\!\unit[1500]{mm}$ and $D\!=\!\unit[250]{mm}$ a spatial resolution of $>\unit[60]{lp/mm}$ was measured for an extended object, exceeding the requirements. The measurement was limited by the pixel size of the camera used to evaluate the resulting image -- corresponding simulations show resolutions better than \unit[100]{lp/mm} for a \unit[1]{cm$^2$} object and reasonable alignment and surface precisions.

\begin{figure}[tbp]
  \centering
  \includegraphics[width=7.5cm]{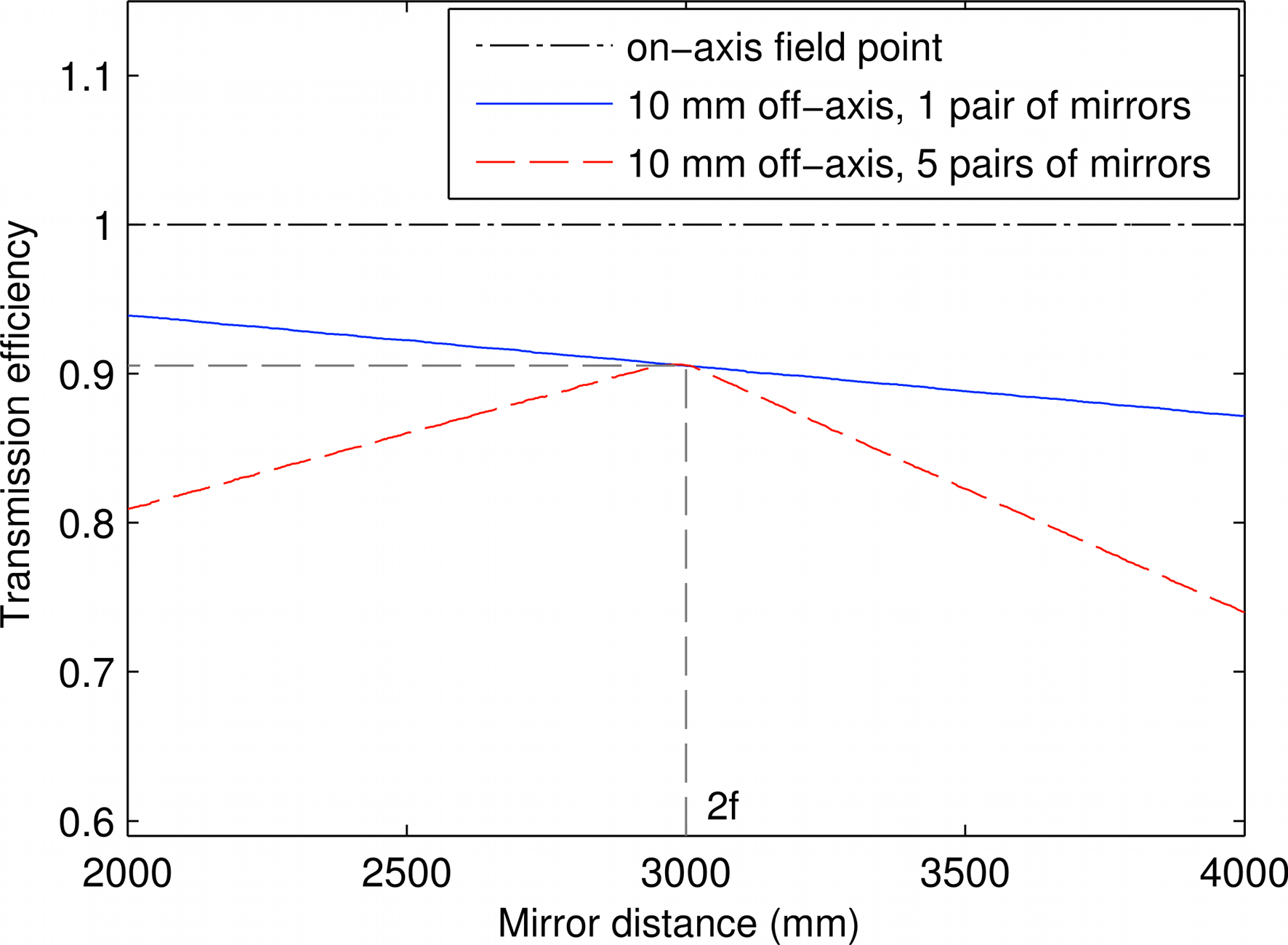}
  \caption{Simulated transmission efficiency for on- and off-axis field points imaged by the system in figure~\ref{fig:parabs_linear} ($f\!=\!\unit[1500]{mm}$, $D\!=\!\unit[250]{mm}$), but varying the mirror distance. See text for further explanation.}
  \label{fig:parabs_lightloss}
\end{figure}

\begin{figure}[tbp]
  \centering
  \includegraphics[width=7.5cm]{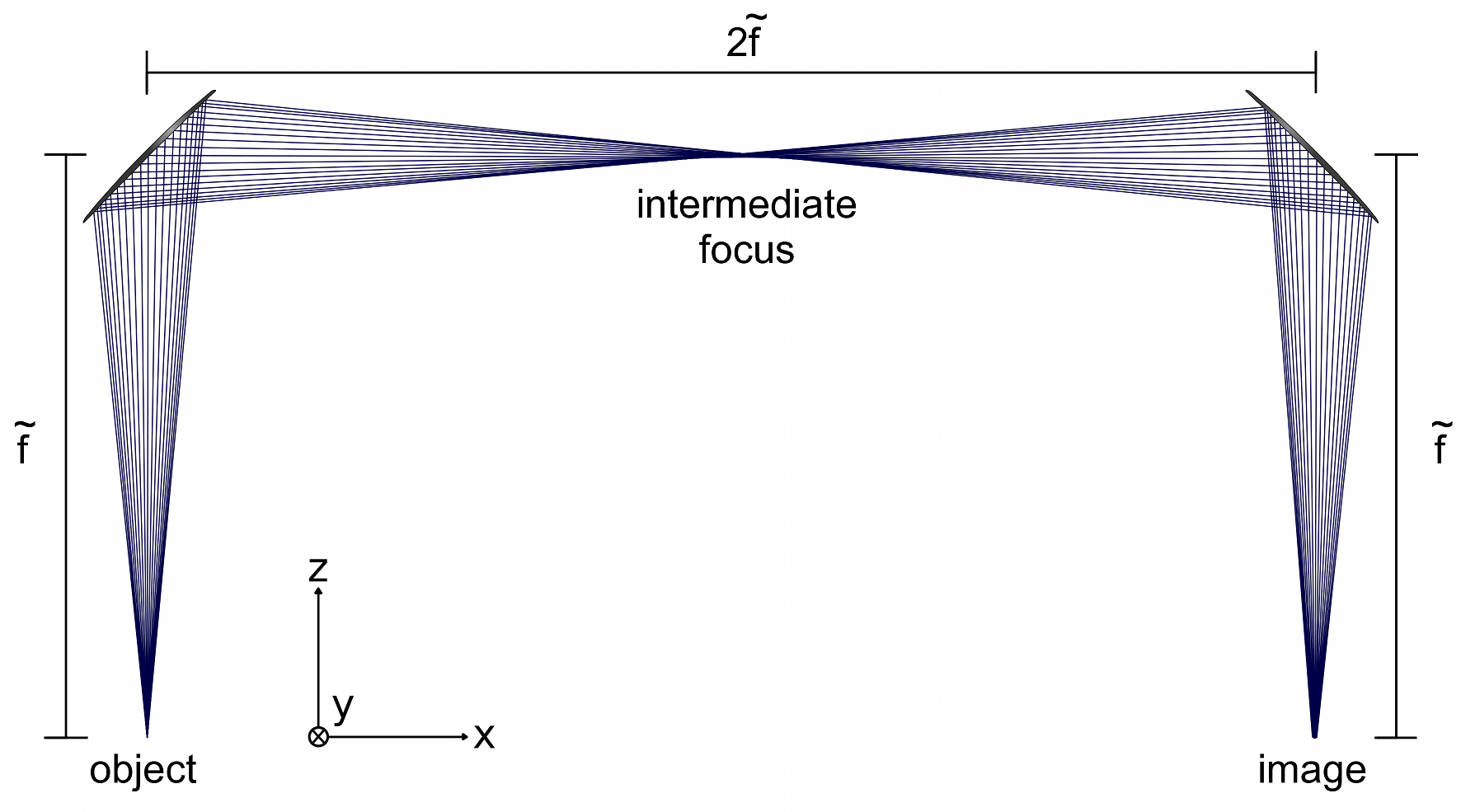}
  \caption{In this setup, on-axis parabolic mirrors have been tilted around the $y$-axis by 45\degree. The image formation works differently in the two directions spanning the image plane: The intermediate focus only occurs in the $x$-$z$-projection, while in the $x$-$y$-projection the rays are propagating parallel to each other between the mirrors. The mirror diameters are exaggerated in this figure.}
  \label{fig:parabs_tilted_axial}
\end{figure}

Due to the mirror spacing of $2f$, the object and image planes are located at the same position, resulting in a length efficiency of $0$. No satisfactory modification toward an improved length efficiency was found:
\begin{itemize}
\item  Introducing plane mirrors will either severely reduce the transmission efficiency (due to obstruction) or not lead to a sufficient improvement of the length efficiency. 
\item Choosing a mirror spacing other than $2f$ will introduce an angular magnification, which in turn leads to light-losses. Figure~\ref{fig:parabs_lightloss} shows the simulated transmission efficiency for systems as shown in figure~\ref{fig:parabs_linear} for varying mirror distances (solid line), and for a hypothetical sequence of five such systems (dashed line). Assuming that the first mirror surface is fully illuminated, off-axis field points are always transported with light-losses. For a distance of $2f\!=\!\unit[3000]{mm}$, the length efficiency is zero, but when the distance differs from $2f$, the light-losses are increased. For short mirror distances this only becomes apparent in a sequence of systems.
\item Tilting the parabolic mirrors will rapidly decrease the spatial resolution and makes the system geometry less practical. An interesting exception are tilts of 45\degree, such as for the system shown in figure~\ref{fig:parabs_tilted_axial}.
\end{itemize}

\begin{figure}[tbp]
  \centering
  \includegraphics[width=7.5cm]{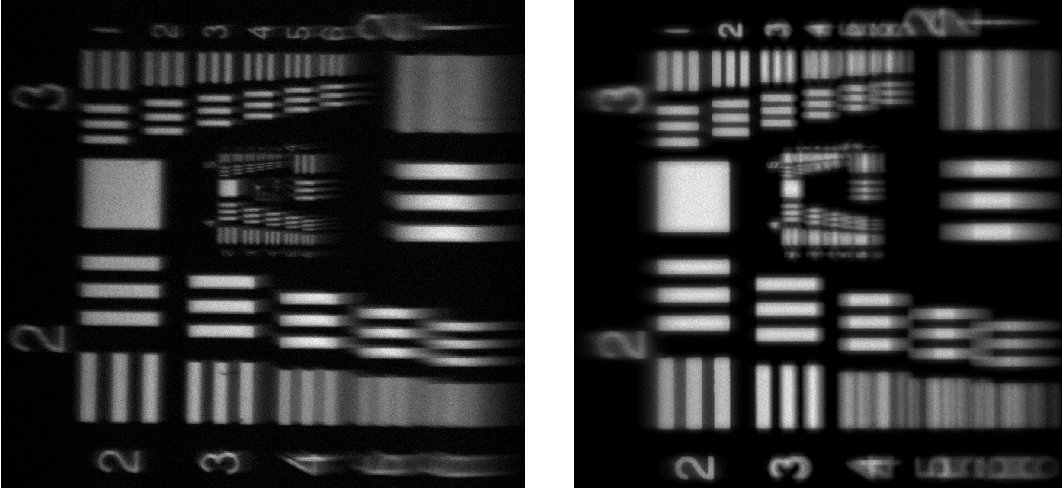}
 \caption{Imaging with two tilted on-axis parabolic mirrors (figure~\ref{fig:parabs_tilted_axial}). \textbf{Left:} Image of a measured $\unit[4]{mm}\times\unit[4]{mm}$ object using two mirrors with an effective aperture of $\tilde f/D\!=\!9.3$. Near the center of the image, a resolution of $\ge\unit[30]{lp/mm}$ was measured (limited by the CCD-camera). Toward the sides of the image, the resolution decays quickly. \textbf{Right:} Corresponding simulation, showing a qualitatively similar behavior, with the resolution degrading toward the sides of the image.}
  \label{fig:on_axis_grid}
\end{figure}

Since parabolic mirrors are readily available, and 45\degree tilts facilitate adapting the system to the building, we studied such setups in detail. The tilt leads to an effective mirror shape (as perceived by a fan of rays extended only in one plane) which is parabolic in one direction and ellipsoid in the orthogonal direction. This results in a curious beam shape between the mirrors: The beam becomes parallel in the first projection and focused in the other, with the ``semi-focus'' located halfway between the two mirrors. Despite this unusual and obviously not radially symmetrical geometry, the system can be used for imaging, but it fails to image extended objects with the required spatial resolution for reasonable apertures, and also has an insufficient temporal resolution on the order of $\unit[1]{ps/m}$. Figure~\ref{fig:on_axis_grid} shows the imaging quality reached with such a system in measurement (left) and simulation (right) for a setup with $\tilde f/D=9.3$, where $\tilde f\!=\!\sqrt2f$ is the effective focal length of the tilted parabolic mirror. If two such systems are arranged in a sequence such that the semi-foci appear in the same plane, the aberrations partially cancel and the imaging quality is improved in simulations, but this was not observed in practice (after spending a reasonable amount of time on the alignment).

\subsection{Off-Axis Parabolic Mirrors}
\label{sec:off-axis-parab-mirr}
\begin{figure}[tbp]
  \centering
  \includegraphics[width=7.5cm]{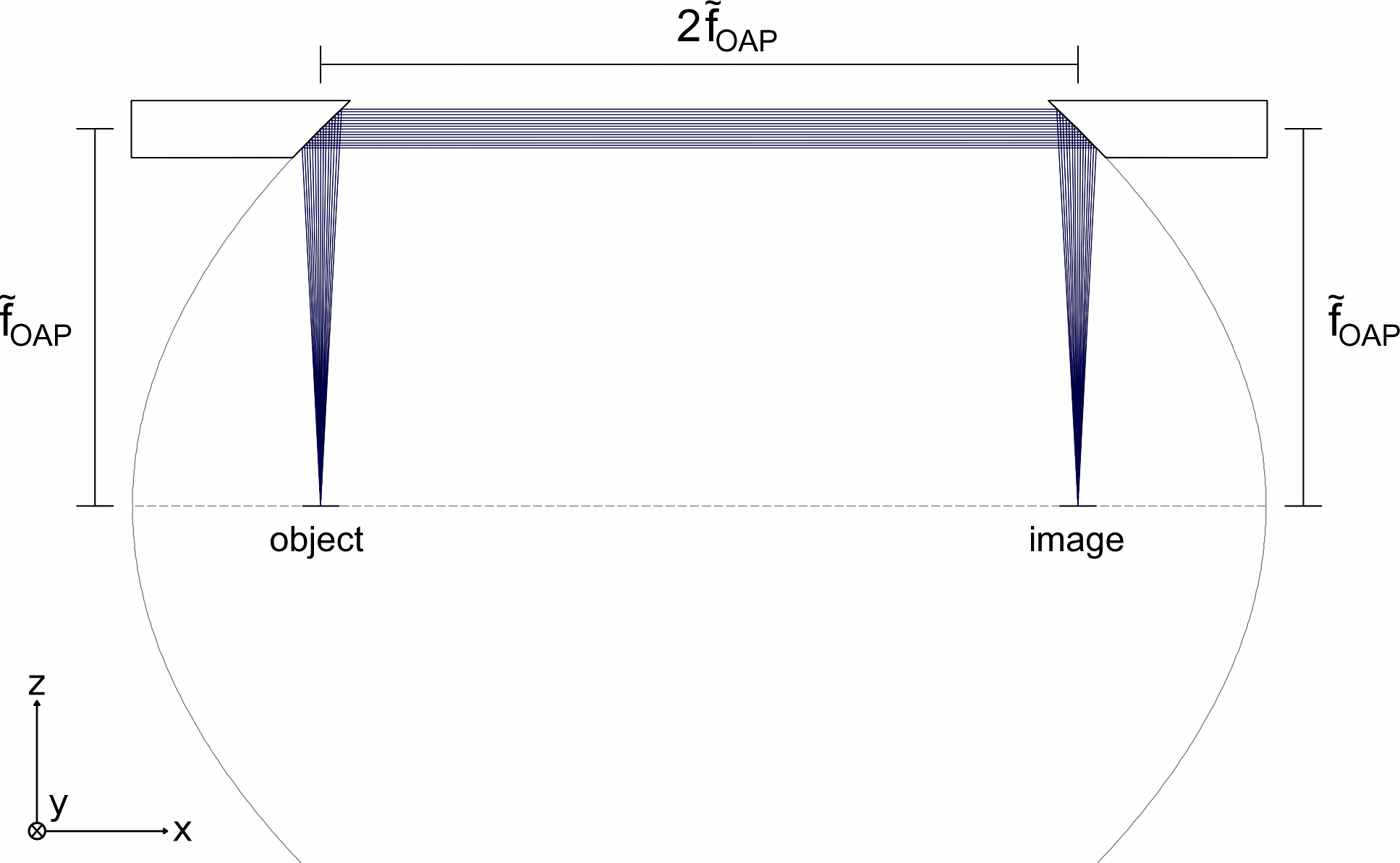}
  \caption{Setup of two OAP mirrors. At first glance it appears similar to the setup in figure~\ref{fig:parabs_tilted_axial}, but it has very different properties. Notice also the absence of an intermediate focus. The outlines of the complete parabolas and their rotational symmetry axis are shown for clarity. }
  \label{fig:oap_4fscheme}
\end{figure}

To a certain extent, the setup of tilted parabolic mirrors can be interpreted as having its object far away from the optical axis. In a different approach, an object on the optical axis is used in combination with an off-axis entrance pupil, resulting in a mirror shape often referred to as off-axis parabolic (OAP) mirror. An off-axis pupil version of the initial (i.e. not tilted) system of two parabolic mirrors is shown in figure~\ref{fig:oap_4fscheme}. Despite the strong deviation between the symmetry axis and the actual light path, such systems perform well. The length efficiency is 50\%, but can be improved by adding plane mirrors. Length efficiencies above 90\% are easily attainable for reasonably sized OAP mirrors. Raytracing simulations of perfectly aligned systems show spatial resolutions close to the diffraction limit and near perfect temporal resolutions on the order of a few femtoseconds. This remains true for extended objects, provided their diameter is small compared to the focal length (up to $\sim5$ percent).

Monte Carlo tolerancing simulations of an OAP system with $\tilde f_{OAP}\!=\!\unit[1000]{mm}$ and $D\!=\!\unit[150]{mm}$ were run in order to understand the requirements on surface quality and alignment precision. The producer of the mirrors (Kugler\footnote{Kugler GmbH, Heiligenberger Str. 100, 88682 Salem, Germany}) specified a surface shape deviation of $<\!\!\unit[1]{\um}$ Peak-to-Valley\footnote{Deviation from best-matching paraboloid, with $\tilde f_{OAP}\!\approx\!\unit[1000]{mm}$.} and a roughness of $<\!\!\unit[10]{nm}$. These tolerancing simulations showed that this surface quality is sufficient, but also revealed that the alignment tolerances for the rotational degrees of freedom are tight. The most critical parameter is the relative tilt between the two mirrors about the $y$-axis (see figure~\ref{fig:oap_4fscheme}), which must not exceed \unit[0.5]{mrad} to fulfill the system specifications. A set of mirrors was ordered, and the alignment procedure and image quality were tested. To facilitate the alignment, simulations of the effects of individual misalignments were made, and supports that allow adjustment of all 3 translational and 3 rotational axes were constructed.
The measured spatial resolution of~\unit[25]{lp/mm} for an extended object exceeds the requirements for the main section.
The MC simulation results suggest that the measured resolution is limited due to the achievable alignment precision. The temporal resolution was not measured in the lab, as no suitable light-source was available. The improvement can be estimated by integrating the OAP setup as a subsystem into the existing OTL, resulting in a hybrid system of lenses and mirrors.

\section{Hybrid System of Lenses and Mirrors}
\label{sec:hybrid-system-lenses}
The encouraging results reached in the calculations and the first tests in the laboratory with OAP mirrors at unit magnification are foreseen to be tested in the existing OTL based mainly on achromatic lenses~\citep{NIM10_JR}. There are three telescopes in the existing system which are imaging at unit magnification (see figure~\ref{fig:overall_system_layout}). They are proposed to be replaced by telescopes of OAP mirrors. The resulting hybrid system consists of reflective optics for unit magnification and of refractive optics for other magnifications.

The four achromatic lenses of the main section of the current OTL are to be replaced by two telescopes of two parabolic mirrors each with an effective focal length of \unit[1900]{mm}. The input branch of the OTL for the measurement of the longitudinal phase space distribution in the high energy section (branch 4 in figure~\ref{fig:overall_system_layout}) also contains a telescope of two achromatic lenses ($f\!=\!~\unit[1000]{mm}$) with unit magnification, which also could be replaced by a telescope of two parabolic mirrors with an effective focal length of~\unit[1250]{mm}. The new focal lengths were chosen such that little repositioning of the residual existing optics is required, although this resulted in a length efficiency of 0.8 for the telescope in branch 4, slightly below the specifications. Plane mirrors are utilized to have the new optical axis close to its original course, thus barely impacting the length efficiency.

\begin{figure}[t]
\center
\includegraphics[width=7.5cm]{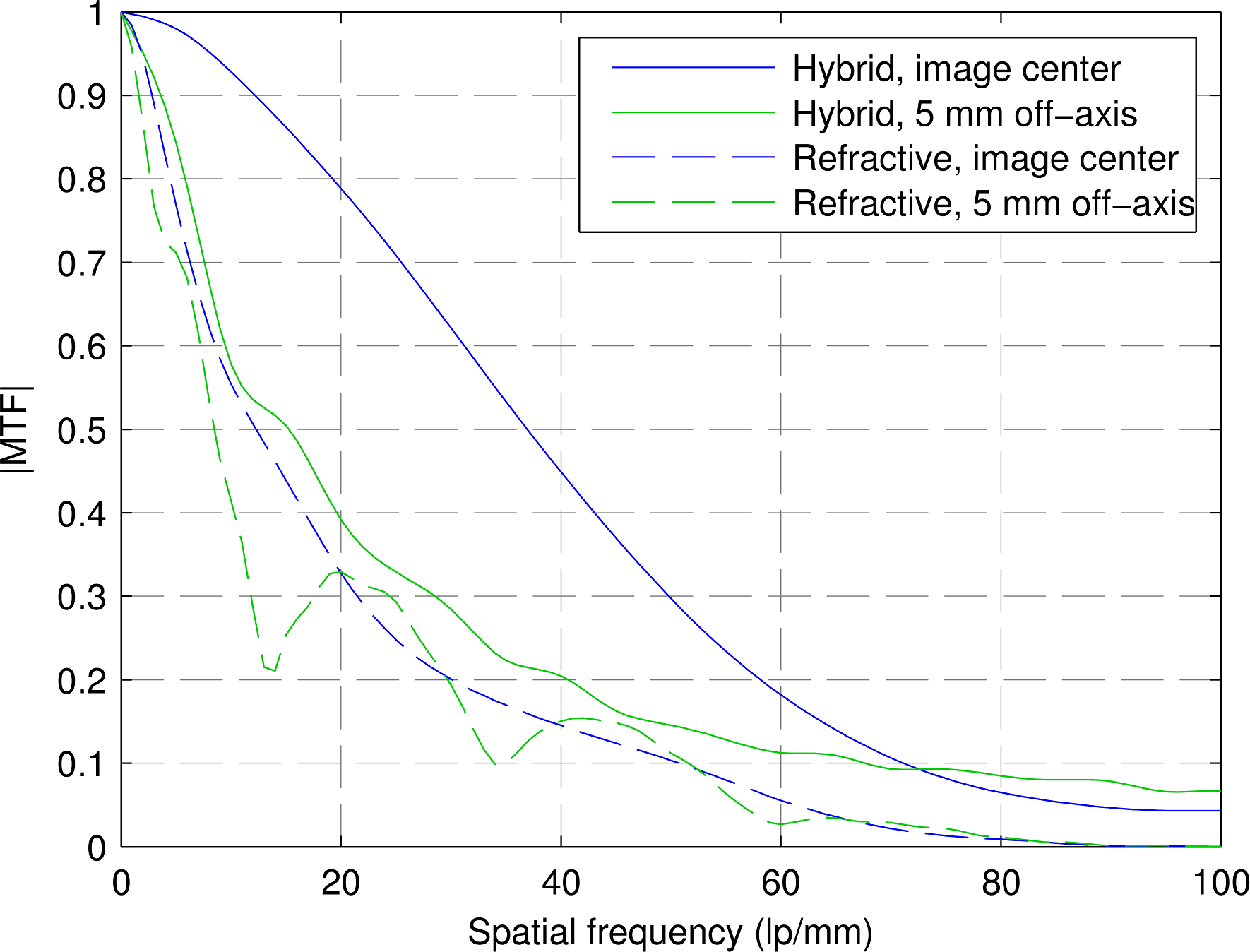}
\caption{The MTF of the existing OTL (for the low energy dispersive arm, branch 1) compared to the MTF of the hybrid system, for field points at the image center and \unit[5]{mm} off-axis (geometric mean of tangential and sagittal resolution).}
\label{fig:mtf_hybrid}
\end{figure}

The performance of this hybrid system was analyzed in further simulations. Figure~\ref{fig:mtf_hybrid} shows the simulated MTF as a function of the spatial frequency in~\unit{lp/mm}. The MTF is shown for the existing OTL and for the hybrid system, in both cases for the input branch of the low energy dispersive arm (branch 1 in table~\ref{tab:input_sections}).

The transverse resolution of the hybrid system is slightly better than for the current OTL. This means the OAP mirrors provide a higher transverse resolution than the lenses they replace. The remaining lenses limit the achievable resolution.

\begin{figure}[t]
\center
\includegraphics[width=7.5cm]{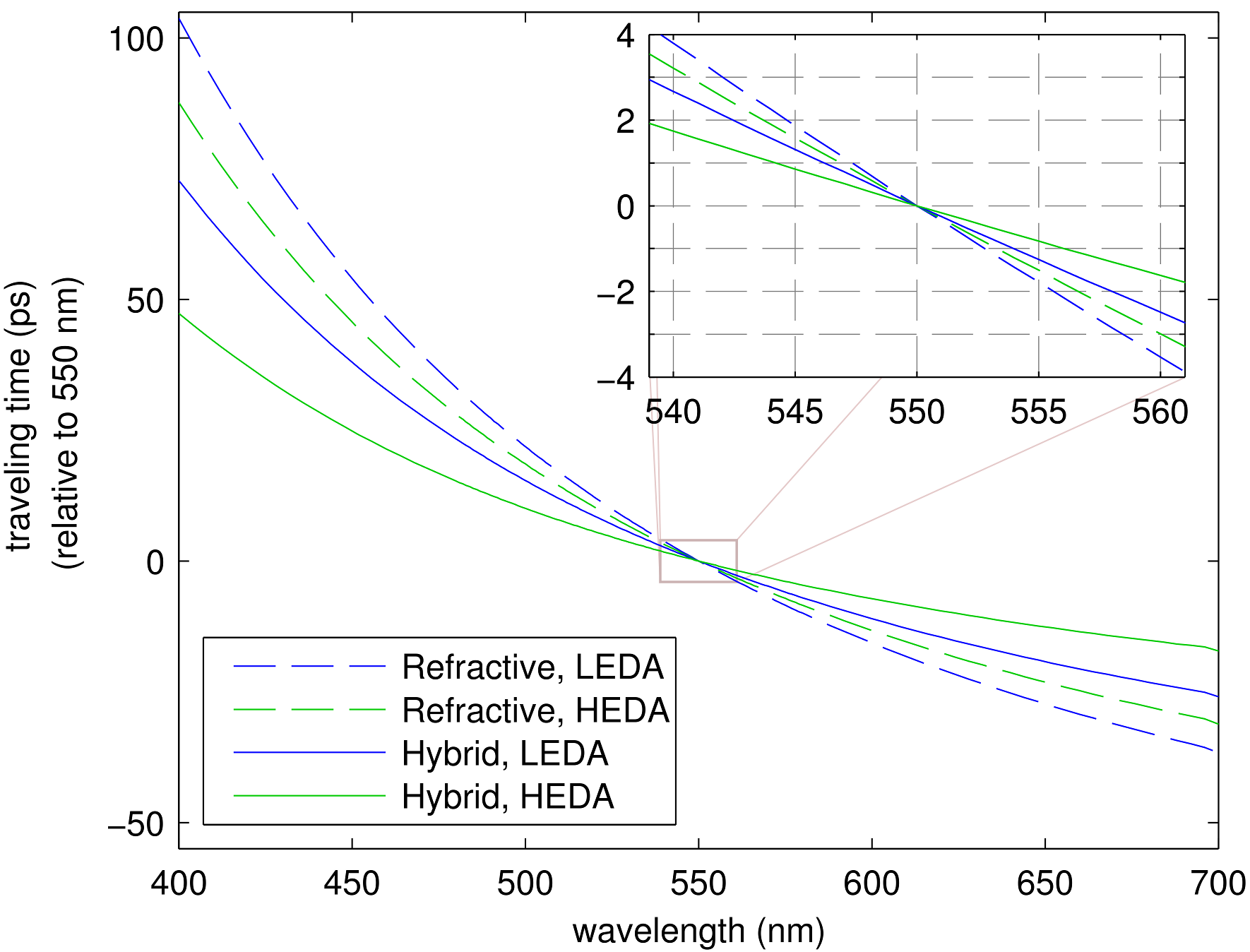}
\caption{Comparison of the calculated traveling time of the Cherenkov light pulses at the streak camera as a function of the wavelength between existing and planned system, for the input branches of the low and high energy dispersive arms (LEDA and HEDA). \unit[550]{nm} was chosen as reference wavelength because the most intense light output occurs at this wavelength.}
\label{fig:dt_hybrid}
\end{figure}

In figure~\ref{fig:dt_hybrid}, the traveling times of the light pulses to the streak camera are shown as a function of the wavelength, using the traveling time of the \unit[550]{nm} component as a reference. They have been computed both for the existing and the planned OTL, for the input branches of the low and the high energy dispersive arms (branches 1 \& 4 in figure~\ref{fig:overall_system_layout}). Measurements and simulations of the current OTL are in very good agreement~\citep{NIM10_JR}.

The temporal resolution is expected to improve when using the hybrid system. The difference in traveling times between \unit[545]{nm} and \unit[555]{nm} is reduced from \unit[3.6]{ps} to \unit[2.6]{ps} ($-28\%$) for the low energy dispersive arm and from \unit[3.1]{ps} to \unit[1.6]{ps} ($-48\%$) for the high energy dispersive arm. Alternatively, an interference filter of a higher bandwidth can be used to increase the signal intensity. For example, using an interference filter of \unit[20]{nm} bandwidth (FWHM), a traveling time difference of \unit[3.4]{ps} for the high energy dispersive arm is expected for the hybrid system, compared to \unit[3.1]{ps} with a \unit[10]{nm} bandwidth filter for the current system. In this way, the signal intensity at the streak camera can be doubled without losing much temporal resolution.

\section{Schemes for Magnification and Demagnification}
\label{sec:schem-magn-demagn}
Any two individually well-corrected telescopes of different focal lengths $f_1$ and $f_2$ can be used in sequence to obtain a system with a magnification $m\!=\!\frac{f_2}{f_1}$. However, for the classical telescopes mentioned before, no combination was found that satisfied $m\!=\!\pm0.1$ as well as $\Delta t<\unit[1]{ps}$. Using two OAP mirrors of different focal lengths is not possible without losing the symmetry of the system, leading to an insufficient resolution. 

A doublet mirror has sufficient degrees of freedom to allow correction of four Seidel aberrations~\cite{BS_Optik}, for example: 2 surface curvature radii $r_i$, 2 conic constants $\delta_i$, and the mirror spacing $d$. The distances and curvature radii can be scaled by a constant, which will also affect the mirror diameters. This scaling factor has only a minor influence on the spatial resolution and should be chosen based on source properties and available space. The remaining four parameters can be used for the corrections. A corrected physical image can be realized for any desired magnification $|m|$, limited however to $m<0$.

\begin{figure}[b!]
  \centering
  \includegraphics[width=7.5cm]{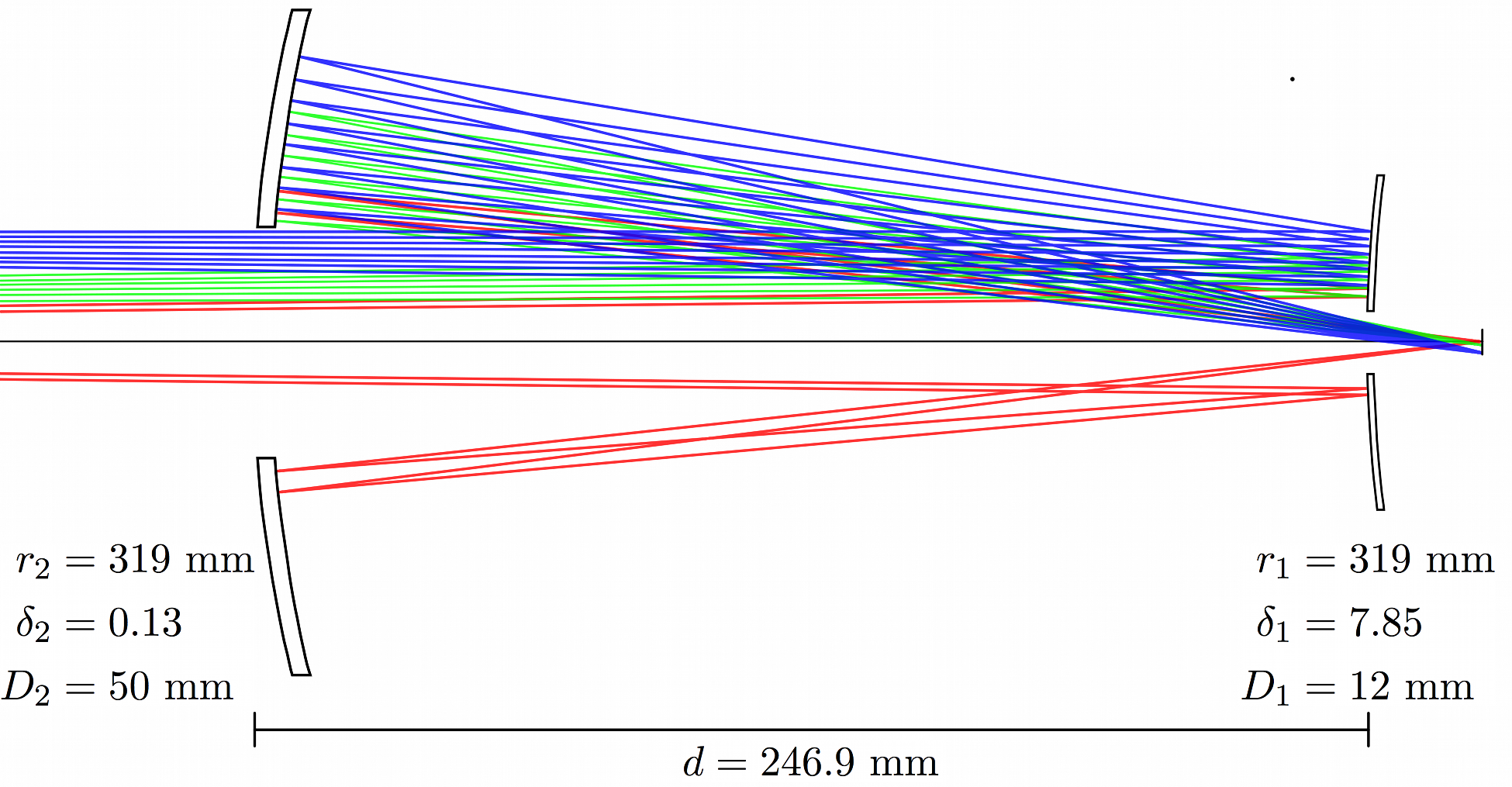}
  \caption{Layout of a doublet system with $m\!=\!0.1$ satisfying the requirements of the matching section, showing the radii $r_i$ and conic constants $\delta_i$ for primary and secondary mirror. The diameters of the inner apertures $D_i$ of the mirrors control the amount of light collected from different field points, but a fully homogeneous illumination is not possible.}
  \label{fig:doublet}
\end{figure}

Figure~\ref{fig:doublet} shows the layout and the parameters of a system with $m\!=\!0.1$ and~\unit[1000]{mm} distance from object to image plane. The simulated spatial resolution of an \mbox{$54\!\!\times\!\!\unit[54]{mm^2}$} object exceeds \unit[50]{lp/mm} everywhere in the image. The simulated temporal resolution of this system is about~\unit[1]{ps}, which is at the upper limit of the acceptable range, since the initial magnification will further worsen the temporal resolution by a similar amount. However, if the object size is reduced to \mbox{$54\!\!\times\!\!\unit[1]{mm^2}$}, corresponding to the streak camera slit size of \mbox{$5.4\!\!\times\!\!\unit[0.1]{mm^2}$}, the temporal resolution is improved to about~\unit[100]{fs}. The system has an inhomogeneous light collection: The ratio between the most and least illuminated field points is roughly 2:1. This property strongly depends on the selected mirror opening diameters and the details of the Cherenkov model used in the simulation. Further optimization and tolerancing studies are required.

\section{Summary and Outlook}
\label{sec:summary-outlook}
The current optical system used for streak camera measurements suffers from dispersion and radiation damage to the lenses. An investigation toward a reflective optical system was started to overcome these problems, with an emphasis on systems working at unit magnification. A selection of mirror systems based on classical telescopes or parabolic mirrors were simulated. Different systems of on-axis or off-axis parabolic mirrors were studied experimentally. The analysis showed that off-axis parabolic mirrors performed very well in terms of spatial resolution and temporal resolution, while imposing acceptable requirements on surface quality and alignment precision. So far, the temporal resolution was only studied in simulations, but will be measured after installation of a hybrid system of lenses and mirrors. If this proof of principle is successful, other parts of the optical system can be replaced in the future, to further increase the temporal resolution and transmission efficiency.
The work on input branches and final matching optics toward a fully reflective optical system is being continued.

\section*{Acknowledgments}
The authors would like to thank the whole PITZ team, in particular Frank Stephan for support and discussions, and Gerald Koss for the design of the OAP mirror supports.

\bibliographystyle{elsarticle/elsarticle-num}

\end{document}